\documentclass[12pt]{article} 
\usepackage[sectionbib]{natbib}
\usepackage{array,epsfig,fancyheadings,rotating}
\usepackage[]{hyperref}  
\usepackage{sectsty, secdot}
\sectionfont{\fontsize{12}{14pt plus.8pt minus .6pt}\selectfont}
\renewcommand{\theequation}{\thesection\arabic{equation}}
\subsectionfont{\fontsize{12}{14pt plus.8pt minus .6pt}\selectfont}

\usepackage{amsmath}
\usepackage{amssymb}
\usepackage{amsfonts}
\usepackage{multirow}
\usepackage{amsthm}
\usepackage{qtree}
\usepackage{booktabs}

\theoremstyle{definition}

\newtheorem{thm}{Theorem}
\newtheorem{lemma}{Lemma}

\newtheorem{defin}{Definition}
\newcommand{\vx}{{\bf x}}

\newcommand{\vX}{{\bf X}}

\newcommand{\vtheta}{\mbox{\boldmath $\theta$}}

\newcommand{\vbeta}{\mbox{\boldmath $\beta$}}

\newcommand{\vZero}{{\bf 0}}

\newcommand{\bqa}{\begin{eqnarray*}}
\newcommand{\eqa}{\end{eqnarray*}}
\newcommand{\bqan}{\begin{eqnarray}}
\newcommand{\eqan}{\end{eqnarray}}
\newcommand{\bit}{\begin{itemize}}
\newcommand{\eit}{\end{itemize}}
\newcommand{\ben}{\begin{enumerate}}
\newcommand{\een}{\end{enumerate}}
\newcommand{\beq}{\begin{equation}}
\newcommand{\eeq}{\end{equation}}
\newcommand{\bdes}{\begin{description}}
\newcommand{\edes}{\end{description}}

\begin{document}


\renewcommand{\baselinestretch}{2}

\markright{ \hbox{\footnotesize\rm Statistica Sinica
}\hfill\\[-13pt]
\hbox{\footnotesize\rm
}\hfill }

\markboth{\hfill{\footnotesize\rm Adriano Zanin Zambom AND Seonjin Kim AND Nancy Lopes Garcia} \hfill}
{\hfill {\footnotesize\rm VLMC with Exogenous Covariates} \hfill}

\renewcommand{\thefootnote}{}
$\ $\par


\fontsize{12}{14pt plus.8pt minus .6pt}\selectfont \vspace{0.8pc}
\centerline{\large\bf Variable Length Markov Chain}
\vspace{2pt} \centerline{\large\bf with Exogenous Covariates}
\vspace{.4cm} \centerline{Adriano Zanin Zambom, Seonjin Kim, and Nancy Lopes Garcia} \vspace{.4cm} \centerline{\it
California State University Northridge, Miami University Ohio,} \centerline{\it and State University of Campinas} \vspace{.55cm} \fontsize{9}{11.5pt plus.8pt minus
.6pt}\selectfont


\begin{quotation}
\noindent {\it Abstract:}
Markov Chains with variable length are useful stochastic models for data compression that avoid the curse of dimensionality faced by that full Markov Chains. In this paper we introduce a Variable Length Markov Chain whose transition probabilities depend not only on the state history but also on exogenous covariates through a logistic model. The goal of the proposed procedure is to obtain the context of the process, that is, the history of the process that is relevant for predicting the next state, together with the estimated coefficients corresponding to the significant exogenous variables. We show that the proposed method is consistent in the sense that the probability that the estimated context and the coefficients 
are equal to the true data generating mechanism tend to 1 as the sample size increases. Simulations suggest that, when covariates do contribute for the transition probability, the proposed procedure outperforms variable length Markov Chains that do not consider covariates while yielding comparable results when covariates are not present.

\vspace{9pt}
\par
\end{quotation}\par

\def\thefigure{\arabic{figure}}
\def\thetable{\arabic{table}}

\renewcommand{\theequation}{\thesection.\arabic{equation}}

\fontsize{12}{14pt plus.8pt minus .6pt}\selectfont

\setcounter{section}{1} 
\setcounter{equation}{0} 

\lhead[\footnotesize\thepage\fancyplain{}\leftmark]{}\rhead[]{\fancyplain{}\rightmark\footnotesize\thepage}

\noindent {\bf 1. Introduction}

Estimating higher order Markov Chains can be challenging even when the process takes values in a finite state space. The difficulty arises from the fact that the number of parameters to be estimated increases exponentially with the order of the chain.
Stochastic chains with memory of variable length constitutes a parsimonious class of higher order Markov Chains. The idea is that for each past, only a finite suffix of the past, called {\sl context}, is enough to predict the next symbol.  These contexts can have different lengths depending on the past itself and can be described by a rooted tree. These chains were first introduced by \cite{Rissanen1983} as a tool for data compression.  For a fixed set of contexts, estimation of the 
transition probabilities can be easily achieved. The problems lies into estimating the contexts from the available data. In his seminal 1983 paper, Rissanen introduces the {\it Context algorithm}, which aims at estimating a flexible class of tree structured models by lumping together irrelevant states in the history of the process. When the order of the tree is bounded regardless of the size of the sample, the proposed algorithm estimates in a consistent way both the length of the context as well as the associated transition probability. These Markov Chains with variable order, or variable length, were further modified with a statistical perspective by \cite{BuhlmannWyner1999}, who showed that the context algorithm is consistent even when the order of the chain is allowed to grow with the sample size.

Classical Markov Chain methods and their variable length versions have been well developed theoretically, however, failing to include in the model available time-dependent covariates that actually affect the transition probabilities may lead to incorrect estimation of the structure of the context tree and its parameters. \cite{MuenzRubinstein1985} provide a remarkable insight about modeling the transition probabilities of  two-state Markov chains with exogenous covariates using logistic regression. They develop maximum likelihood estimation of the first order Markov chain parameters with a time-invariant covariate, that is, the covariates are fixed characteristics. Several possible extensions are briefly discussed, which include the first order Markov chain with time-dependent covariates and the second order Markov chain with time-invariant covariates. After this pioneer work, there has been a growing interest in studying Markov chains with exogenous covariates (see for example \cite{MuenzRubinstein1985}, \cite{Azzalini1994}, \cite{BarrantesEtAl1995}, \cite{CookNg1997}, \cite{VermuntEtAl1999}, \cite{AalenEtAl2001}, \cite{Heagerty2002}, \cite{IslamChowdhury2006}, \cite{RubinEtAl2017}, \cite{SirdariIslam2018}). On the other hand,  formal inference regarding Markov Chains  with exogenous covariates has not been carefully addressed. First, the transition probabilities in the literature modeled through logistic regression are restrictive and do not make full use of predictive significance of possibly time-variant exogenous covariates that are observed concomitantly with the process. To the best of our knowledge,  all work in  the literature allows only one observation of covariates, which is either the initial, a time-invariant value, or the most recently measured value. 
Second, inference about the order of the Markov Chain with exogenous covariates  has not been developed. The order is closely related to the exogenous covariates because the predictive significance of past covariates may determine the transition probabilities.
For instance, if a second order Markov model is employed, then it is natural to use in the logistic regression the values of the covariates observed at to the two most recent time points.
Furthermore, in order to develop a more sophisticated Markov model that can address these issues, formal inference needs to be established when lumping together irrelevant past states and deciding which past covariates should be included in the model.

In response to this gap in literature, we propose the {\it beta-context} model with variable length, which allows the structure of the Markov Chain to depend on exogenous covariates throughout time.  
For Markov Chains with variable length, the concept of context is defined as the values of the infinite history of the process that are relevant for predicting the next state.
Based on the idea in \cite{BuhlmannWyner1999}, we develop a novel backward selection algorithm 
for Markov chains with exogenous covariates with variable length by considering both the history of binary response states and the predictive significance of past covariates. A fitted context model provides a context tree, where every branch corresponds to a history of response states which is associated with parameters that define the transition probability in terms of a logistic regression. Therefore, the transition probability depends on both past states and values of past covariates. More precisely speaking, the model assumes that the transition probabilities are defined by the logistic regression model, whose parameters are linked to the state history.

As the name of the model, the predictive significance of past covariates on the next state of the process plays a critical role in constructing the beta-context model with variable length. Note that  the {\it beta} represents the parameters in logistic regression.  
In the proposed backward selection procedure, the predictive significance of the covariates in the past most state is evaluated first by the maximum likelihood based inference. If it turns out that the covariates have a significant predictive ability, then the state is kept in the context tree. Otherwise, the state is re-evaluated without the covariates to check whether it could be merged with its sibling, which is the state that has the same parent node in the context tree. This pruning avoids the curse of dimensionality by removing nodes that are not significant. We prove that the proposed context algorithm is consistent for estimating the true tree as well as estimating the coefficients in logistic regression as the type 1 error rate converges to zero at a certain rate. 

The remainder of the paper is organized as follows. Section \ref{sec:model1} introduces the notation, the model and the beta-context algorithm, including the theory for its consistency. Section \ref{sec:simul} presents simulations results for the finite sample performance of the algorithm, while Section \ref{sec:data} exhibits the application to a real dataset.

\section{The Beta-Context Model with Variable Length}  \label{sec:model1}
\indent \indent 
Consider a stochastic process $(Y_t)_{t\in \mathbb{Z}}$ taking values on a finite state space $\mathcal{Y}$. Denote by $y_i^j = y_j, y_{j-1}, \ldots, y_i (i < j, i, j \in \mathbb{Z} \cup \{-\infty, \infty\}, y_j \in \mathcal{Y})$ a string written in reverse time representing the states visited by the process from time $i$ to time $j$.  Suppose that the transition probabilities of the process may depend on the previous states through a set of parameters and $d$ exogenous covariates. Similarly to \cite{Rissanen1983} and \cite{BuhlmannWyner1999}, we consider processes whose context lengths may depend on the actual values of the state history. Denote the $\ell$-th covariate observed at time $t$ by $X_{t\ell}$, its observed value by $x_{t\ell}$, $\ell = 1, \ldots, d$, and  let $\vX_t= (X_{t1}, \ldots, X_{td})$, $\vx_t= (x_{t1}, \ldots, x_{td})$.
 Further define
$$
\vX_{i}^j = (\vX_{j},\vX_{j-1},\dots,\vX_{i})\quad \mbox{and} \quad \vx_{i}^j = (\vx_{j},\vx_{j-1},\dots,\vx_{i})
$$
as the vector of covariates and the vector of their observed values, respectively, from time $i$ to time $j$.

The {\it beta-context} of a stochastic process $(Y_t)_{t\in \mathbb{Z}}$, which will be detailed in this section, is characterized by the dependence of both state history and covariate values. For simplicity of presentation and notation ease, in Subsections \ref{sec:binary}--\ref{sec:theory} we restrict the state space to $\mathcal{Y} = \{0, 1\}$. More specifically, the transition probabilities of the Markov Chain are defined by a logistic regression model with linear effects of the exogenous covariates,  and parameters that are dependent on the state history.
 Still under the binary scenario, the beta-context algorithm is described in Subsection \ref{sec:algorithm} and the consistency of the estimators obtained with the proposed algorithm is investigated in Subsection \ref{sec:theory}.
A generalization of the model, algorithm and theory for processes on finite state spaces is discussed in Subsection \ref{sec:multinomial}.

\subsection{Binary State Space} \label{sec:binary}

The {\it beta-context} function $c(\cdot)$ of a stochastic process $(Y_t)_{t\in \mathbb{Z}}$ taking values in the state space $\mathcal{Y} = \{0, 1\}$ is defined as follows.

\begin{defin} \label{Def1} Let $(Y_t)_{t\in \mathbb{Z}}$ be a stationary process with values $Y_t \in \mathcal{Y}$ 
and $(\vX_t)_{t\in \mathbb{Z}}$ a family of  $d$-dimensional  vectors of  exogenous covariates with $ \vX_t \in \mathcal{X} \subseteq \mathbb{R}^d$. Denote by $c: \mathcal{Y}^{\infty} \to \mathcal{Y}^{\infty}$ a (projection) function which maps 
\bqa
&&c: y_{-\infty}^0 \to y_{-\ell+1}^0, \text{ where }  \ell  \text{ is defined by} \\
&&\hspace{-.5cm}\ell = \min\Big\{k: P(Y_1 = 1|Y_{-\infty}^0 = y_{-\infty}^0, \vX_{-\infty}^0 = \vx_{-\infty}^0)\\
&& \hspace{.1cm}= P_{\vtheta}(Y_1 = 1|Y_{-k+1}^0 = y_{-k+1}^0, \vX_{-h+1}^0 = \vx_{-h+1}^0) \text{ for  all } \vx_{-\infty}^0 \text{ and }  k \geq h \Big\}\\
&&(\ell = 0 \text{ corresponds to independence})
\eqa
and, letting $u := y_{-\ell+1}^0$, 
\bqan\label{Prob_def}
&&P_{\vtheta}(Y_1 = 1|Y_{-\ell+1}^0 = y_{-\ell+1}^0, \vX_{-h+1}^0 = \vx_{-h+1}^0)\nonumber\\
&&\hspace{2cm} = \frac{\exp(\alpha^u + \sum_{t=-h+1}^{0} \sum_{\ell=1}^ d x_{t\ell} \beta^u_{t\ell})}{1 + \exp( \alpha^u + \sum_{t=-h+1}^{0} \sum_{\ell=1}^ d x_{t\ell} \beta^u_{t\ell})},
\eqan
   for $h \leq \ell$. We denote by  $\vtheta :=  \vtheta^{u} = (\alpha^u, \vbeta^u) = \left(\alpha^u,  \vbeta^u_0, \ldots, \vbeta_{(-h+1)}^u\right)$, the vector of coefficients associated with the context (past states) $u = y^0_{-\ell+1}$ for transitioning into the state $1$, where $\vbeta_{t}^u = ( \beta_{t1}^u, \ldots, \beta_{td}^u)$ is the vector of coefficients corresponding to the $d$ exogenous covariates at time $t$, $t = 0, \ldots, -h+1$. 
Then, $c(\cdot)$ is called the {\it beta-context function} for any $t \in \mathbb{Z}$, and $c(y_{-\infty}^{t-1})$ is called the {\it beta-context} for the transition at time $t$  with associated parameter vector $\vtheta^u$. 
\end{defin}

Note that each parameter vector $\vbeta^u$ depends on the outcome of the function $c(\cdot)$, i.e. $u$, and can be composed of zero or non-zero elements. The length of $\vbeta^u$, which represents the number of steps where covariate values have significant contribution to the model, denoted here by $h$ ($h := h_u$), can be thought of the past most index of the elements of $\left(\vbeta^u_0, \ldots, \vbeta_{(-\ell+1)}^u\right)$ that is non-zero, where $\ell$ is the length of its associated context $u = y_{-\ell+1}^0$. More specifically, for a context $u$, if $\vbeta^u = \left(\vbeta^u_0, \ldots, \vbeta_{(-h+1)}^u, 0, \ldots, 0\right)$ with $h \leq \ell$, then the length of $\vbeta^u$ is $h$.
We formalize the length of this type of vector with the following definition.

\begin{defin} \label{Def_Length} Consider a parameter vector $\vbeta^u = \left(\vbeta^u_0, \ldots, \vbeta_{(-\ell+1)}^u\right)$ associated with a beta-context $c(\cdot) = u$. The length of $\vbeta^u$ is defined as
\bqa
h := h_u =  |\vbeta^u| = 1-\min_{j = 0, \ldots, -\ell+1}\{j: \vbeta_j^u \neq 0\}.
\eqa
If $\vbeta_0^u = \ldots = \vbeta_{-\ell+1}^u = 0$, then $|\vbeta^u| = 0$.
\end{defin}

It is important to notice that from its definition, the length of the beta-context $\ell = |c(\cdot)|$ and the length $h$ of the associated parameter vector $\vtheta$ are variable and implicitly dependent on the observed state history $u = y_{-\ell+1}^0$.
 The transition probability function in (\ref{Prob_def}) determines what values of the infinite history of $y_{-\infty}^0$ are relevant along with the observed values of the exogenous covariates. 
The proposed model allows $\ell \geq h$, which includes scenarios where the transition probability can depend on a longer history of the state space transitions but possibly only recent covariate history.
As such, the VLMC defined in \cite{BuhlmannWyner1999} without exogenous variables is a special case of Definition \ref{Def1}, where the direct influence of the context $u$ in the transition probability can be seen as determined by $\alpha^u$, so that setting all coefficients $\vbeta^u$ to zero reduces the proposed model to the VLMC.

\begin{defin} Let $(Y_t)_{t\in \mathbb{Z}}$ be a stationary process with values $Y_t \in \mathcal{Y}, |\mathcal{Y}| < \infty$,  $(\vX_t )_{t\in \mathbb{Z}}$ a family of $d$-dimensional vectors of  exogenous covariates, $\vX_t \in \mathcal{X} \subseteq \mathbb{R}^d$, and $c(\cdot)$ the corresponding beta-context function defined in Definition \ref{Def1}. Let $0 \leq \eta \leq \infty$ be the smallest integer such that 
\bqa
|c(y_{-\infty}^0)| = \ell(y_{-\infty}^0) \leq \eta, \text{ for all } y_{-\infty}^0 \in \mathcal{Y}^{\infty}.
\eqa
Then $c(\cdot)$ is called a beta-context function of order $\eta$, and $(Y_t)_{t\in \mathbb{Z}}$ is called a stationary beta-context model of order $\eta$. 
\end{defin}

The order of the context function determines the maximum number of steps in any context which contains relevant information about the future.
Following \cite{BuhlmannWyner1999}'s formulation of a context tree based on the context function, we define a beta-context tree for the beta-context function $c(\cdot)$ of order $\eta$.  Because the transition probabilities are determined by the regression model, each context in the tree is accompanied with an associated vector of parameters.

\begin{defin} Let $c(\cdot)$ be a beta-context function of a stationary beta-context model of order $\eta$. 
The $(|\mathcal{Y}|$-ary) beta-context rooted tree $\tau$  
is defined as
\bqa
\tau &:=& \tau_c = \{u: u = c(y_{-\eta+1}^0), y_{-\eta+1}^0 \in \mathcal{Y}^{\eta} \}
\eqa
with an associated parameter tree
\bqa
\tau_\theta &=&   \{(u,\vtheta^{u}): u \in \tau\}
\eqa
where $\vtheta^{u}$ is defined in Definition \ref{Def1}.
\end{defin}

The tree $\tau_\theta$ contains all the information about the data generating system, that is, the contexts $u$ that determine which states visited by the chain are significant to the transition probability and the parameters $\vtheta$ associated with each of these contexts.

\subsection{The Beta-Context Algorithm}\label{sec:algorithm}

\indent \indent The goal of the beta-context algorithm is to, based on data composed by $n$ state transitions $Y_1^n$ and the respective values of the  covariate vector $\vX_1^n$, estimate the underlying beta-context function $c(\cdot)$ and the transition probability parameters $\vtheta^u$. 
Define 
\bqa
n_v := N(v) = \sum_{t=1}^n I(Y_{t}^{t+|v|-1} = v), \;\; v \in \mathcal{Y}^\infty,
\eqa
which represents the number of occurrences of the string $v$ in the sequence $Y_1^n$. 
In Definition \ref{Def_Sibling}  we define the concept of siblings and parents in context trees, which will be used in the description of the algorithm and in the theoretical results of Section \ref{sec:theory}.

\begin{defin} \label{Def_Sibling} Let $u_1 = uw \in \tau$ and $u_2 = uw' \in \tau$, for $w, w' \in \mathcal{Y}$ and $u \in \mathcal{Y}^\infty$, be two contexts whose last nodes are the only ones that differ. Then $u_1$ and $u_2$ are called siblings in $\tau$ and this relationship is denoted by $u_1\wr u_2$. In addition, $u$ is called the parent of $u_1$ and $u_2$.
\end{defin}

The proposed algorithm starts with a maximal tree and prunes the final nodes iteratively if they are not significant. Differently from the context trees in \cite{BuhlmannWyner1999} and \cite{Rissanen1983}, here the pruning of nodes is based not only on the significance of the context, but also on the possible influence of the exogenous covariates through the coefficients $\vbeta$. Hence, the likelihood of the data $Y_1^n$ conditionally on $\vX_1^n = \vx_1^n$ based on the model in Definition \ref{Def1} is
\bqan\label{Likelihood}
&& L(\tau_\theta;Y_1^n = y_1^n, \vX_1^n = \vx_1^n) = P(Y_1^n = y_1^n|\vX_1^n = \vx_1^n, \tau_\theta)\nonumber\\
 &=& P(Y_1 = y_1)P(Y_2 = y_2|Y_1 = y_1, \vX_1 = \vx_1, \tau_\theta)\nonumber\\
 && \hspace{1cm}\ldots P(Y_n = y_n|Y_1^{n-1} = y_1^{n-1},\vX_1^{n-1} = \vx_1^{n-1}, \tau_\theta),
\eqan
where each probability follows from (\ref{Prob_def}).

\indent \indent For the context $u = y_{-k+1}^0 = y_0, y_{-1}, \ldots, y_{-k+1}$ with $k$ steps into the past there are $1+dk$ parameters to be estimated: 1 corresponding to $\alpha^u$ and $dk$ corresponding to the $d$ exogenous covariates at each of the $k$ steps. In order to avoid poor performance of the likelihood ratio test in the algorithm, a minimum number $s \geq 1$ of observations per parameter to be estimated is required, so that the number of observations deemed necessary to estimate the $1+dk$ parameters for each of the contexts $u$  is $s(1+dk)$. The algorithm is as follows.

{\bf Step 1.} Given the data $(Y_1^n, \vX_1^n)$, 
fit a maximal $|\mathcal{Y}|$-ary beta-context rooted tree, that is, find the beta-context $c_{max}(\cdot)$ 
where 
\bqa
&&\hspace{-.8cm}\tau_{\max}=\{ u_1 = y_{-k+1}^0, u_2: u_1\wr u_2, N(u_1)\geq s(1+dk), N(u_2)\geq s(1+dk)\},\\
&&\tau_{max} \supseteq \tau, \text{ where } u = y_{-k+1}^0 \in \tau \text{ implies } N(y_{-k+1}^0) \geq s(1+dk).
\eqa
Set the  initial beta-context tree as $\tau^{(0)} = \tau_{max}$, denote its order by $r$, and let the associated parameter tree be $\tau^{(0)}_{\theta}$.  Note that  $\tau^{(0)} = \tau_{max}$ may not be a full tree, that is, the final nodes contexts may not all have the same length.
Compute $\hat\tau^{(0)}_{\theta}$, the associated estimated parameter tree, where the parameters are estimated via maximizing the likelihood $L(\tau_{\theta}^{(0)}|y_1^n, \vx_1^n)$.

{\bf Step 2.} For each context $u \in \tau^{(0)}$ of length $r$, use the LRT to test the significance of the parameter vector corresponding to the covariates from the past most node, i.e., 
\bqa
H_0^u: \vbeta_{-r+1}^u = 0.
\eqa
To do so, let 
\bqan\label{Lambda_u}
\lambda^{u}_{-r+1} &=& -2 \Big[\log L\left(\tilde\tau_{\theta}^u\Big|y_1^n, \vx_1^n\right) - \log L\left(\hat\tau_{\theta}^{(0)}\Big|y_1^n, \vx_1^n\right)\Big]
\eqan
 be the deviance statistic for testing $H_0^u$, where $L(\cdot)$ is the likelihood defined in (\ref{Likelihood}). The estimators 
 \bqan\label{eq.tau_tilde}
 \tilde\tau_{\theta}^u = \{(w,\tilde\vtheta^w): w \in \tau^{(0)}, w \neq u\}\cup\{ (u,(\tilde\alpha^{u}, \tilde\vbeta^{u}_0, \tilde\vbeta^u_{-1},\ldots, \tilde\vbeta^{u}_{-r+2}, 0))\}
 \eqan
   are estimated maximizing the likelihood under the null hypothesis $H_0^u$.
Define the p-value $\pi^{u}_{-r+1} = 1 - \Psi_{d}(\lambda^{u}_{-r+1})$, where $\Psi_{d}(\cdot)$ is the cumulative distribution function of a $\chi^2$ random variable with $d$ degrees of freedom.
If $\pi^{u}_{-r+1} > \gamma_n$, for a chosen level $\gamma_n$, update $\hat\tau^{(0)}_{\theta}$ with $\tilde\tau_{\theta}^u$.

{\bf Step 3.} With respect to the tests performed in {\bf Step 2}:

a) If neither  $H_0^{u_1}$ nor $H_0^{u_2}$, for $u_1$ and $u_2$ siblings in $\tau_{(0)}$, was rejected then 
test whether these siblings' past most nodes can be dropped: let 
\bqan\label{Test_Siblings}
\lambda^{u_1u_2} &=& -2 \Big[\log L\left(\tilde\tau_{\theta}^{u_1u_2}\Big|y_1^n, \vx_1^n\right) - \log L\left(\hat\tau_{\theta}^{(0)}\Big|y_1^n, \vx_1^n\right)\Big],
\eqan
where $\tilde\tau_{\theta}^{u_1u_2} = \{(w,\tilde\vtheta^w): w \in \tau, w \neq u_1, u_2\}\cup\{ (u,\tilde\vtheta^u): u \text{ is parent of } u_1 \text{ and } u_2\}$ is estimated under the null hypothesis that the sibling nodes are both replaced by their parent, and consequently reducing the parameters from the two vectors $\vtheta^{u_1}$ and $\vtheta^{u_2}$, which have been reduced to size $\mathbb{R}^{1+d(r-1)}$ in Step 2,  to the parameters in $\vtheta^u \in \mathbb{R}^{1+d(r-1)}$.  If $\pi^{u_1u_2} = 1 - \Psi_{d(r-1)+1}(\lambda^{u_1u_2}) \geq \gamma_n$, replace the siblings $u_1 = uw$ and $u_2 = uw'$ with their parent $u$, updating $\tau^{(0)}$, and updating $\hat\tau_\theta^{(0)}$ with $\tilde\tau_{\theta}^{u_1u_2}$.

b) If at least one of $H_0^{u_1}$ and $H_0^{u_2}$, for $u_1$ and $u_2$ siblings in $\tau^{(0)}$, was rejected, both $u_1$ and $u_2$ remain in the tree. Then sequentially test to prune the past most parameters in $\vbeta^{u_1}$ and/or $\vbeta^{u_2}$, whichever had its hypothesis not rejected in Step 2, up to the root as follows.  Let $u_1 = y_{-r+2}^0$ so that $\vbeta^{u_1} = (\vbeta_0^{u_1}, \ldots, \vbeta_{-r+2}^{u_1})$ 
Let $j = -r+2$, the past most index of $\vbeta^{u_1}$, and test $H^{u_1}_{0j}: \vbeta^{u_1}_{j}= 0$ using the test statistic $\lambda^{u_1}_j$ defined as in (\ref{Lambda_u}). If $\pi^{u_1}_j = 1 - \Psi_{d}(\lambda^{u_1}_j) > \gamma_n$ set $\vbeta^{u_1} = (\vbeta_0^{u_1}, \ldots, \vbeta_{-r+3}^{u_1})$, and set $j = -r+3$. Otherwise stop and keep all $\vbeta^{u_1} = (\vbeta_0^{u_1}, \ldots, \vbeta_{-r+2}^{u_1})$. Repeat this test sequentially for $j = -r+2, \ldots, 0$, until $\pi^{u_1}_j < \gamma_n$ or when $|\vbeta^{u_1}| = 0$. Repeat this process for 
$\vbeta^{u_2}$.

At the end of {\bf Step 3}, both covariate influence (measured by the parameter coefficients), and context tree at level $r$ have been possibly pruned, generating a possibly smaller context tree $\tau^{(1)} \subseteq \tau^{(0)}$ and its corresponding updated parameter tree $\tau^{(1)}_{\hat\theta}$.

{\bf Step 4.} Repeat Steps 2 and 3 with the updated trees $\tau^{(1)}$ and $\tau^{(1)}_{\hat\theta}$ for contexts of length $r-1, r-2, \ldots, 1$. The exception is that, if $u_1$ and $u_2$ are siblings and one of them has children, there is no context pruning (both nodes are kept) but the pruning of covariate parameters is done sequentially until the root as in {\bf Step 3} part b).

Denote this pruned beta-context tree by $\hat{\tau}_n$ with associated parameter tree $\hat{\tau}_\theta$ and corresponding beta-context function $\hat{c}(\cdot)$.

\subsection{Consistency of the beta-Context Algorithm}\label{sec:theory}

In this section we show that the beta-context algorithm described in Section \ref{sec:algorithm} will produce estimates for the beta-context tree as well for the parameters of the regression and the transition probabilities which are strongly consistent. In order to prove these results we will need a bound for the test statistic used for the regression parameter under the alternative hypothesis which will be given in Lemma \ref{Main.Lemma} and the following assumptions concerning the cut-off parameter $\gamma_n$ and the size of the maximal tree:

C1: $\gamma_n \to 0$ such that $n\gamma_n = o(1)$.

C2: $\gamma_n \to 0$ such that $(1/n)\log(1/\gamma_n) = o(1)$.

C3: The order of the initial maximal tree $\tau_{max}$ is $r = O(log(n))$.

\begin{lemma} \label{Main.Lemma} Let $\lambda^{u}_{-r+1}$ be the test statistic for the hypothesis $H_0^u: \vbeta_{-r+1}^u = 0$ vs $H_a^u: \vbeta_{-r+1}^u \neq 0$ as defined in (\ref{Lambda_u}). Then under the alternative $H_a^u$
\bqa
\lambda^{u}_{-r+1}  &\geq& O_p(n).
\eqa
\end{lemma}

Let $\hat{\tau}_{n}$ and $\hat{\tau}_{\theta_{n}} =  \{(u,\hat{\vtheta}^{u}_n): u \in \hat{\tau}_{n}\}$ be the estimated beta-context tree and its associated parameter tree computed using the beta-context algorithm described in Section \ref{sec:algorithm}. Theorem \ref{Main.Thm} established that $\hat{\tau}_{n}$  converges almost surely to the the true data generating mechanism denoted by the tree $\tau$. In addition, given a consistent estimator $\hat{\tau}_{n}$ of $\tau$, $\hat{\tau}_{\theta_{n}}$ is strongly consistent for $\tau_\theta$ in the sense that both the estimated parameter tree and its associated parameters converge almost surely to their population counterparts. 

\begin{thm} \label{Main.Thm} Assume the beta-context tree $\tau$ has finite order. Then, under conditions C1-C3, there exists an integer-valued random variable $N$ with $P(N < \infty) = 1$ such that

a) $\hat{\tau}_{n} = \tau$ $\forall n \geq N$ with probability 1,

b) $|\hat{\vtheta}^u_n| = |\vtheta^u|$ $\forall u \in \tau,$ $\forall n \geq N$ with probability 1,

c) $\hat{\vtheta}^u_n \to \vtheta^u$ $\forall u \in \tau,$ as $n \to \infty$ with probability 1.
\end{thm}

\subsection{Finite State Spaces}\label{sec:multinomial}

For stochastic processes that take values on a finite state space $\mathcal{Y} = \{1, \ldots, p\}$, the {\it beta-context} function can be defined similarly to that of the binary case. The main difference is that the transition probabilities into each of the $p$ states are built with distinct sets of parameters, which compose a multinomial regression model. As a consequence, the pruning steps of the {\it beta-context} algorithm must handle the multiple final nodes in the tree.
Definition \ref{DefGeneral} generalizes the {\it beta-context} function $c(\cdot)$ in Definition \ref{Def1} to the multivariate case.

\begin{defin} \label{DefGeneral} Let $(Y_t)_{t\in \mathbb{Z}}$ be a stationary process with values $Y_t \in \mathcal{Y}$ 
and $(\vX_t)_{t\in \mathbb{Z}}$ a family of  $d$-dimensional  vectors of  exogenous covariates with $ \vX_t \in \mathcal{X} \subseteq \mathbb{R}^d$. Denote by $c: \mathcal{Y}^{\infty} \to \mathcal{Y}^{\infty}$ a (projection) function which maps 
\bqa
&&c: y_{-\infty}^0 \to y_{-\ell+1}^0, \text{ where }  \ell \text{ is defined by} \\
&&\ell = \min\Big\{k: P(Y_1 = y_1|Y_{-\infty}^0 = y_{-\infty}^0, \vX_{-\infty}^0 = \vx_{-\infty}^0)\\
&& \hspace{2cm}= P_{\vtheta}(Y_1 = y_1|Y_{-k+1}^0 = y_{-k+1}^0, \vX_{-h}^0 = \vx_{-h+1}^0), \text{ for all } \vx_{-\infty}^0,\\
&& \hspace{5cm}  k \geq h, \text{ and for all }  y_1 \in \mathcal{Y}\}\\
&&(\ell = 0 \text{ corresponds to independence})
\eqa
and, letting $u := y_{-\ell+1}^0$  
\bqa
&&P_{\vtheta}(Y_1 = y_1|Y_{-\ell+1}^0 = y_{-\ell+1}^0, \vX_{-h+1}^0 = \vx_{-h+1}^0)\\
&& = \frac{\exp(\alpha_{y_1}^u + \sum_{t=-h+1}^{0} \sum_{\ell=1}^ d x_{l\ell} \beta^u_{y_1,t\ell})}{\sum_{j = 1}^p\exp( \alpha_{j}^u + \sum_{t=-h+1}^{0} \sum_{\ell=1}^ d x_{l\ell} \beta^u_{j,t\ell})} \text{ for all } y_1 \in \mathcal{Y},
\eqa
   for $h \leq \ell$. We denote by $\vtheta := \vtheta^{u} = (\alpha^u_{1}, \ldots, \alpha^u_{p}, \vbeta^u_{1}, \ldots, \vbeta^u_{p})$, with $\vbeta_{j}^u = \left(\vbeta_{j,0}^u, \ldots, \vbeta_{j,(-h+1)}^u\right)$, the vector of coefficients associated with the past states $u = y^0_{-\ell+1}$ for transitioning into state $j \in \mathcal{Y}$, where $\vbeta_{j,t}^u = ( \beta_{j,t1}^u, \ldots, \beta_{j,t d}^u)$ is the vector of coefficients corresponding to the $d$ covariates at time $t = 0, \ldots, -h+1$. 
Then, $c(\cdot)$ is called the beta-context function for any $t \in \mathbb{Z}$, and $c(y_{-\infty}^0)$ is called the beta-context for the transition at time 1 with associated parameter vector $\vtheta^u$. 
\end{defin}
For identifiability reasons, for each set $\left((\alpha_{1}^u, \vbeta_{1}^u), \ldots, (\alpha_{p}^u,\vbeta_{p}^u)\right)$ we restrict one $(\alpha_{j}^u,\vbeta_{j}^u)$ to be equal to a $id+1$ vector of zeros, which is called the baseline category. For the special case where $p = |\mathcal{Y}| = 2$, this corresponds to the logit link in the logistic regression addressed in Subsections \ref{sec:binary}-\ref{sec:theory}.

A natural and direct generalization of the {\it beta-context} algorithm to handle the multiple siblings at the pruning steps can be formulated in the following way.
First, Steps 1, 2, and 4 remain the same, only now the contexts may be composed transitions into any state in the state space $\mathcal{Y} = \{1, \ldots, p\}$. Step 3 can be modified as follows.

{\bf Step 3.} With respect to the tests performed in {\bf Step 2}:

a) If no $H_0^{u_j}, j = 1, \ldots, p$, for $u_1, \ldots, u_p$ siblings in $\tau_{(0)}$, was rejected then 
test whether all these siblings' final nodes can be dropped together: let 
\bqan\label{Test_Siblings}
\lambda^{u_{1\ldots p}} &=& -2 \Big[\log L\left(\tilde\tau_{\theta}^{u_{1\ldots p}}\Big|y_1^n, \vx_1^n\right) - \log L\left(\hat\tau_{\theta}^{(0)}\Big|y_1^n, \vx_1^n\right)\Big],
\eqan
where $\tilde\tau_{\theta}^{u_{1\ldots p}} = \{(w,\tilde\vtheta^w): w \in \tau, w \neq u_1,\ldots  u_p\}\cup\{ (u,\tilde\vtheta^u): u \text{ is parent of } u_1,\ldots   u_p\}$ is estimated under the null hypothesis that the sibling nodes are all replaced by their parent, and consequently reducing the parameters from the vectors $\vtheta^{u_1}, \ldots \vtheta^{u_p}$, which have been  each reduced to size $\mathbb{R}^{1+d(r-1)}$ in Step 2,  to the parameters in $\vtheta^u \in \mathbb{R}^{1+d(r-1)}$.  If $\pi^{u_{1\ldots p}} = 1 - \Psi_{(p-1)(d(r-1)+1)}(\lambda^{u_{1\ldots p}}) \geq \gamma_n$, replace the siblings $u_1, \ldots, u_p$ with their parent $u$, updating $\tau^{(0)}$, and updating $\hat\tau_\theta^{(0)}$ with $\tilde\tau_{\theta}^{u_{1\ldots p}}$.

b) If at least one of $H_0^{u_j}, j = 1, \ldots, p$, for $u_1, \ldots, u_p$ siblings in $\tau^{(0)}$, was rejected, all $u_1, \ldots, u_p$ remain in the tree. Then sequentially test to prune the past most parameters in $\vbeta^{u_j}, j = 1, \ldots, p$ which had its hypothesis not rejected in Step 2, up to the root similarly to the case of the binary tree.

This modification of the {\it beta-context} algorithm prunes all siblings together when they are not jointly significant. Showing the consistency of the estimated context function in this case, in the sense of Theorem \ref{Main.Thm}, is trivial since the rates in equations (S1.4) and (S1.2) in the supplementary material become $O(\gamma_np^{r+1})$ instead of $O(\gamma_n2^{r+1})$.

\section{Simulations}\label{sec:simul}

In this section we evaluate the finite sample performance of the proposed beta-context algorithm in three different scenarios. For comparison purposes, the results obtained with \cite{BuhlmannWyner1999}'s context algorithm are also presented.
We generate the data with models from context trees of orders 3 and 4 and varying lengths of the covariate parameter vector $\vbeta^u$ with univariate exogenous covariate. The first model considered is

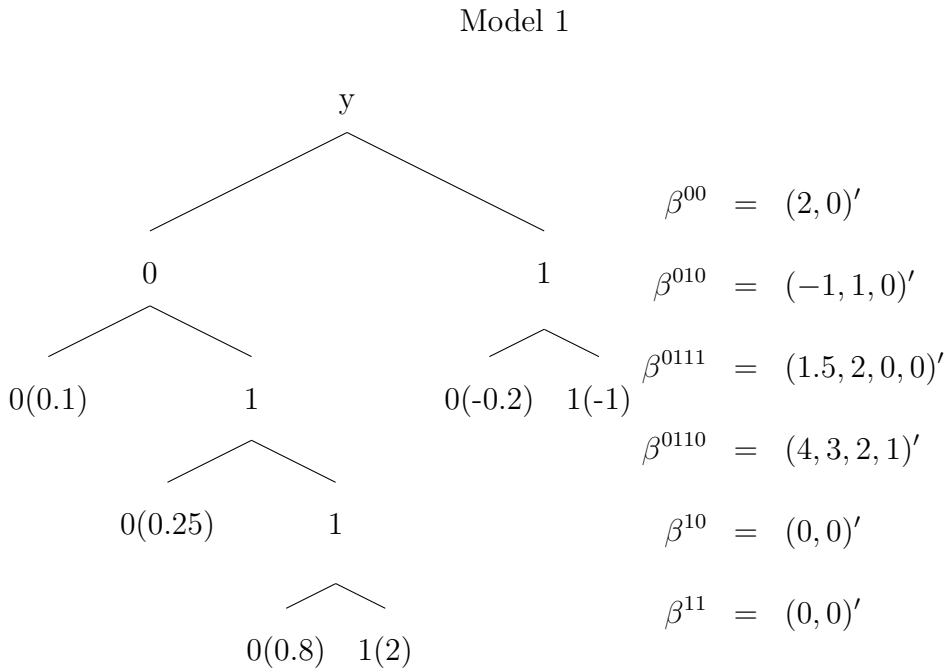
\begin{center}
Model 1
\end{center}
\begin{minipage}{0.45\textwidth}
\Tree[.y [.0 [.0(0.1) ]
                  [.1 [.0(0.25) ]
                       [.1 [.0(0.8) ]
                            [.1(2) ] ]]]
             [.1 [.0(-0.2)  ]
                  [.1(-1) ]]]
\end{minipage}%
\hfill
\begin{minipage}{0.45\textwidth}
\bqa
\beta^{00} &=& (2,0)'\\
\beta^{010} &=& (-1,1,0)'\\
\beta^{0111} &=& (1.5,2,0,0)'\\
\beta^{0110} &=& (4,3,2,1)'\\
\beta^{10} &=& (0,0)'\\
\beta^{11} &=& (0,0)'
\eqa
\end{minipage}

\noindent
where the numbers in parenthesis represent the values of $\alpha^u$ for each context $u$. Note that the length of the coefficient vector $\beta^{0111}$ is 2, which means that only the two observations $x_{n-1}$ and $x_{n-2}$ are relevant in this model for the context $(0,1,1,1)$. 
For example, for the data $(y_1, \ldots, y_5) = (0,1,1,1,0)$ and univariate exogenous variable observations $\vx_1^5 = (x_1, \ldots, x_5)$, the probability that the next observation is a 1 is 

\bqa
P(Y_6 = 1|Y_{1}^5 = (0,1,1,1,0), \vX_{1}^5 = \vx_{1}^5) &=& P(Y_6 = 1|Y_{2}^5 = (0,1,1,1), \vX_{2}^5 = \vx_{2}^5)\\
 &=& \frac{\exp(2 + 1.5x_{5} + 2x_{4})}{1 + \exp( 2 + 1.5x_{5} + 2x_{4})}.
\eqa
This can be seen as if $\ell = 4$ and $h = 2$ in (\ref{Prob_def}).

\vspace{.2cm}
The second model considered is
\begin{center}
Model 2
\end{center}
\begin{minipage}{0.45\textwidth}
\Tree[.y [.0 [.0 [.0(0.5) ] 
                       [.1(0.8) ]]
                  [.1(1) ]]
             [.1 [.0(-0.2)  ]
                  [.1(0.5) ]]]
\end{minipage}%
\hfill
\begin{minipage}{0.45\textwidth}
\bqa
\beta^{000} &=& (3,1,2)'\\
\beta^{001} &=& (1,0,0)'\\
\beta^{01} &=& (-1,-2)'\\
\beta^{10} &=& (-1.2,0)'\\
\beta^{11} &=& (0,0)'.
\eqa
\end{minipage}%

\vspace{.2cm}
The third model, Model 3, is defined exactly as Model 2 except that all $\vbeta^u = \vZero$, so that the performance of the proposed test is evaluated when there is no information about the transition probability in any of the available covariates, that is, under the model in  \cite{BuhlmannWyner1999}. For Models 1, 2, and 3 the exogenous variables $X_i, i = 1, \ldots, n$ were generated independently from a standard Normal distribution. 
For each model, a sample of $n = 1000$ and $n = 2000$ state transitions was generated based on the probability distribution in (\ref{Prob_def}). In order to evaluate the performance of the proposed method we consider several measures on the estimated context function: BIC, AIC, log-likelihood, 
number of parameters $\hat\alpha^u$ (number of final nodes in $\hat\tau$), number of parameters $\hat\vbeta^u$ (total number of coefficients estimated different from 0 in all vectors $\hat\vbeta^u, \forall u$), order of the $\hat\tau$ tree, order of the exogenous covariate (maximum length of $\hat\vbeta^u, \forall u$), number of missing nodes in $\hat\tau$, number of extra nodes $\hat\tau$, $\tau$ tree identified exactly (no missing and no extra nodes in $\hat\tau$), and $\tau_\theta$ tree identified exactly (no longer nor shorter estimated parameter vectors $\hat\vbeta^u, \forall u$). The average of these measures out of 1000 Monte Carlo simulations for Models 1, 2, and 3 with $n = 1000$ are shown in Tables \ref{tab.Model1}, \ref{tab.Model2}, and \ref{tab.Model3} respectively. 
The values of the tuning parameters $\gamma_n$ and $s$ for the proposed beta-context algorithm and "treshold.gen" and "alpha.c" for the VLMC model in \cite{BuhlmannWyner1999} were chosen to minimize the BIC criterium.

\begin{table}[!htbp]
\centering
\footnotesize
\begin{tabular}{lcccccc}
\hline
\hline
Method  & BIC & AIC & logLik &  No. Params $\hat\alpha^u$ & No. Params $\hat\vbeta^u$  \\ 
  \hline
beta-VLMC & 1128.89 & 1093.162 & -539.3008 &  5.395 & 7.28 \\
VLMC &  1345.885 & 1327.599 & -660.0734 &  3.726 & -\\
\cline{2-7}\\
\cline{2-7}
  &  order $\hat\tau$ & order-Covar. & No. Missing $\hat\tau$ & No. Extra $\hat\tau$ & Identical $\tau$ & Identical $\tau_\theta$\\ 
\cline{2-7}
beta-VLMC & 3.908 &  3.874 & 1.282 & 0.072 & 0.371 & 0.041 \\
VLMC & 2.308 & - & 4.632 & 0.084 & 0.017 & -\\
  \hline
  \hline
\end{tabular}
\caption{Simulation results for Model 1 with $n = 1000$ transitions.}
\label{tab.Model1}
\end{table}

\begin{table}[!htbp]
\centering
\footnotesize
\begin{tabular}{lcccccc}
\hline
\hline
Method  & BIC & AIC & logLik & No. Params $\hat\alpha^u$ & No. Params $\hat\vbeta^u$  \\ 
  \hline
beta-VLMC & 1121.727 & 1089.262 & -538.016 &  5.032 & 6.615 \\
VLMC &  1358.512 & 1342.797 & -668.1967 &  3.202 & -\\
\cline{2-7}\\
\cline{2-7}
  &  order $\hat\tau$ & order-Covar. & No. Missing $\hat\tau$ & No. Extra $\hat\tau$ & Identical $\tau$ & Identical $\tau_\theta$\\ 
\cline{2-7}
beta-VLMC & 2.996 & 3.003 & 0.03 & 0.094 & 0.964 &  0.578 \\
VLMC & 2.008 & - & 3.966 & 0.37 & 0.02 & -\\
  \hline
  \hline
\end{tabular}
\caption{Simulation results for Model 2 with $n = 1000$ transitions.}
\label{tab.Model2}
\end{table}

\begin{table}[!htbp]
\centering
\footnotesize
\begin{tabular}{lcccccc}
\hline
\hline
Method  & BIC & AIC & logLik &  No. Params $\hat\alpha^u$ & No. Params $\hat\vbeta^u$  \\ 
  \hline
beta-VLMC & 1275.062 & 1274.974 & -637.4688 &  4.285 & 0.018 \\
VLMC &  1297.462 & 1272.84 & -631.4031 &  5.017 & -\\
\cline{2-7}\\
\cline{2-7}
  &  order $\hat\tau$ & order-Covar. & No. Missing $\hat\tau$ & No. Extra $\hat\tau$ & Identical $\tau$ & Identical $\tau_\theta$\\ 
\cline{2-7}
beta-VLMC & 2.52 & 0.018 & 1.452 & 0.022 & 0.591 & 0.591 \\
VLMC & 3.002 & - & 0.048 & 0.082 & 0.948 & -\\
  \hline
  \hline
\end{tabular}
\caption{Simulation results for Model 3 with $n = 1000$ transitions.}
\label{tab.Model3}
\end{table}

For Models 1 and 2, the proposed beta-context algorithm estimates models with considerable smaller AIC and BIC criteria compared to those of VLMC while for Model 3, the AIC and BIC of the two algorithm are quite comparable. The order of the true data generating tree $\tau$ in Model 1 is 4, suggesting that the vlmc often under-estimates Model 1, while beta-vlmc has an average of estimated trees of order 3.908. Both the proposed procedure and vlmc rarely over-estimate the tree in Model 1, however 1.282 and 4.632 nodes are missed in average respectively for these methods. The proposed method estimates in average a tree $\hat\tau$ that has the exact same nodes as $\tau$ 37.1\% of the time, however, only 4.1\% of the time all coefficients from the significant covariates are identified and none are missing. Although this percentage may seem low, note that it does not mean that the coefficient vector was estimated as 0, but rather that it may have been estimated to have longer or shorter lengths. This still allows $\hat\tau$ to be correctly estimated, as the context is significant with any significant length of the coefficient vector. The results for Model 2 are similar to those in Model 1, except that, with less parameters to estimate, the proposed method achieves 96.4\% accuracy in recovering the true context tree $\tau$, with 57.8\% of the time identifying the exact covariate vectors. On the other hand, for Model 3, the beta-context algorithm recovers the true tree 59.1\% of the time, while vlmc has 94.8\% accuracy. Such a result is expected as Model 3 meets the assumptions of vlmc. Beta-vlmc failure to capture the significant nodes probably comes from the fact that the signal is weak (values of $\alpha^u$ are small) and the proposed method performs more hypothesis tests than the vlmc.

To gain additional insight on the differences between the estimated tree and the original tree, we computed the frequency distribution of the number of missing and extra nodes out of the 1000 simulation runs. Table \ref{tab.FreqModel1} shows the results for Models 1, 2, and 3 when $n = 1000$. These results suggest that for Model 1 the proposed method is missing 2 nodes, or one branch, most of the time, while vlmc is often estimating trees that miss more than 2 nodes, probably due to the failure to detect the signal coming from the exogenous covariate. The frequent extra nodes estimated by vlmc explains its low performance in detecting $\tau$ exactly, only 1.7\%, shown in Table \ref{tab.Model1}. A similar pattern is observed for Model 2. In Model 3, the proposed procedure missed 4 nodes 321 times, which means that the signal was seldom strong enough given the multiple testing adjustment on $\gamma$.

\begin{table}[!htbp]
\centering
\footnotesize
\begin{tabular}{llccccccccccc}
\hline
\hline
Model & Method  & \multicolumn{5}{c}{Missing} & \multicolumn{6}{c}{Extra}  \\ 
\cmidrule(r){3-7}  \cmidrule(r){8-13}
&               & 0 & 2 & 4 & 6 & 8 & 0 & 2 & 4 & 6 & 8 & 10\\
  \hline
\multirow{2}{*}{Model 1} & beta-VLMC & 384 & 591 &   25 & 0 & 0 & 980   & 7 &  10 &   3 & 0  & 0\\
 & VLMC& 27 & 240 & 164 & 528 &  41 & 965  & 30&    4 & 0  & 1  & 0\\
\hline
\multirow{2}{*}{Model 2} & beta-VLMC & 985  & 15 & 0 & 0 & 0 & 979   & 5  &  8  &  7 & 0 &  1 \\
 & VLMC& 159   & 42 &  618 &   19 &  162 & 832 & 153 &   13 &   2 & 0 & 0 \\
\hline
\multirow{2}{*}{Model 3} & beta-VLMC & 595  & 84 &  321 & 0 & 0 & 995   & 1 &   2 &   2  & 0 & 0\\
 & VLMC& 982   & 12 &    6 & 0 & 0 & 965  & 30  &   4 &   1 & 0 & 0\\
\hline
\hline
\end{tabular}
\caption{Frequency distribution of missing and extra nodes $n = 1000$ transitions.}
\label{tab.FreqModel1}
\end{table}

To assess the performance of the proposed method in estimating the vector of coefficients for the exogenous covariates, we computed the mean and standard deviation of the estimators when the non-zero coefficients were correctly identified. Table \ref{tab.BetaModel1} shows the results for Models 1 and 2.  For parameters corresponding to short contexts, for instance $u = (0,0)$, the estimation is accurate and yield a small standard deviation. However, for contexts of longer strings, the estimation over-estimated the parameters with a large standard deviation. This is due to the fact that in $n = 1000$, based on the probabilities of the data generating mechanism, the contexts (0,1,1,0) and (0,1,1,1) did not appear many times in the data generated by Model 1, so that there were not enough observations for an accurate estimation of these 6 parameters. Similarly for the context (0,0,0) in the data generated by Model 2.

\begin{table}[!htbp]
\centering
\footnotesize
\begin{tabular}{lccccccccc}
\hline
\hline
&  \multicolumn{9}{c}{Model 1}\\
\cmidrule(r){2-10}
& $\beta^{00}_1$ & $\beta^{010}_1$ & $\beta^{010}_2$ & $\beta^{0111}_1$ & $\beta^{0111}_2$ & $\beta^{0110}_1$ & $\beta^{0110}_2$ & $\beta^{0110}_3$ & $\beta^{0110}_4$\\
True value & 2 & -1 & 1 & 1.5 & 2 & 4 & 3 & 2 & 1\\
Estimated & 2.03 & -1.06 & 1.04  & 6.91  & 7.05 & 5.38 & 4.11 & 2.79 & 1.39 \\
Std. Dev. & 0.283 & 0.211 & 0.238 & 8.358 & 9.707 & 8.298 & 7.430 & 6.085 & 3.155\\
\hline
&  \multicolumn{9}{c}{Model 2}\\
\cmidrule(r){2-10}
 & $\beta^{000}_1$ & $\beta^{000}_2$ & $\beta^{000}_3$ & $\beta^{001}_1$ & $\beta^{01}_1$ & $\beta^{01}_2$ & $\beta^{10}_1$ \\ 
True value & 3 & 1 & 2 & 1 & -1 & -2 & -1.2\\
Estimated & 5.06 & 1.73 & 3.39 &  1.21 & -1.03 &  -2.03 & -1.21 \\
Std. Dev. & 12.071 & 4.390 & 8.134 & 0.243 & 0.191 & 0.275 & 0.187\\
\hline
\hline
\end{tabular}
\caption{Results for the estimation of the exogenous variable coefficients for $n = 2000$.}
\label{tab.BetaModel1}
\end{table}

To check for the consistency of the proposed method, the simulations of Models 1 and 3 were run with $n = 2000$ state transitions and the results are displayed in Tables \ref{tab.Model1_2000} and \ref{tab.Model3_2000} respectively. Model 2 was omitted as its true tree was identified 96.4\% of the time with $n = 1000$. The results, in view of Tables \ref{tab.Model1} and \ref{tab.Model3}, suggest that as the sample size increases, the proposed algorithm recovers the true data generating mechanism with higher accuracy. This is corroborated by the estimates of the exogenous variables coefficients in Table \ref{tab.BetaModel1_2000}, which were greatly improved when compared to those in Table \ref{tab.BetaModel1}.

\begin{table}[!htbp]
\centering
\footnotesize
\begin{tabular}{lcccccc}
\hline
\hline
Method  & BIC & AIC & logLik &  No. Params $\hat\alpha^u$ & No. Params $\hat\vbeta^u$  \\ 
  \hline
beta-VLMC & 2193.488 & 2144.491 & -1063.498 &  6.068 & 8.748 \\
VLMC &  2674.552 & 2648.132 & -1319.349  & 4.717 & -\\
\cline{2-7}\\
\cline{2-7}
  &  order $\hat\tau$ & order-Covar. & No. Missing $\hat\tau$ & No. Extra $\hat\tau$ & Identical $\tau$ & Identical $\tau_\theta$\\ 
\cline{2-7}
beta-VLMC & 4.08 & 4.054 & 0.116 & 0.252 & 0.889 & 0.429 \\
VLMC & 2.849 & - & 2.688 & 0.122 & 0.049 & -\\
  \hline
  \hline
\end{tabular}
\caption{Simulation results for Model 1 with $n = 2000$ transitions.}
\label{tab.Model1_2000}
\end{table}

\begin{table}[!htbp]
\centering
\footnotesize
\begin{tabular}{lcccccc}
\hline
\hline
Method  & BIC & AIC & logLik &  No. Params $\hat\alpha^u$ & No. Params $\hat\vbeta^u$  \\ 
  \hline
beta-VLMC & 2539.365 & 2539.04 & -1269.462 &  5.023 & 0.058 \\
VLMC &  2573.725 & 2545.458 & -1267.682 &  5.047\\
\cline{2-7}\\
\cline{2-7}
  &  order $\hat\tau$ & order-Covar. & No. Missing $\hat\tau$ & No. Extra $\hat\tau$ & Identical $\tau$ & Identical $\tau_\theta$\\ 
\cline{2-7}
beta-VLMC & 2.992 & 0.058 & 0.048 & 0.094 & 0.959 & 0.958 \\
VLMC & 3.024 & - & 0 & 0.094 & 0.961 & -\\
  \hline
  \hline
\end{tabular}
\caption{Simulation results for Model 3 with $n = 2000$ transitions.}
\label{tab.Model3_2000}
\end{table}

  \begin{table}[!htbp]
\centering
\footnotesize
\begin{tabular}{lccccccccc}
\hline
\hline
&  \multicolumn{9}{c}{Model 1}\\
\cmidrule(r){2-10}
& $\beta^{00}_1$ & $\beta^{010}_1$ & $\beta^{010}_2$ & $\beta^{0111}_1$ & $\beta^{0111}_2$ & $\beta^{0110}_1$ & $\beta^{0110}_2$ & $\beta^{0110}_3$ & $\beta^{0110}_4$\\
True value & 2 & -1 & 1 & 1.5 & 2 & 4 & 3 & 2 & 1\\
Estimated & 2.02 & -1.02 & 1.01 & 2.15 & 2.49 & 4.43 & 3.32 & 2.20 & 1.12 \\
Std. Dev. & 0.196 & 0.167 & 0.159 & 0.651 & 0.934 & 1.032 & 0.794 & 0.574 & 0.390\\
\hline
\hline
\end{tabular}
\caption{Results for the estimation of the exogenous variable coefficients for $n = 2000$.}
\label{tab.BetaModel1_2000}
\end{table}

Overall, the proposed beta-context algorithm outperforms the vlmc in the presence of exogenous covariates, while maintaining competitive results when the transition probabilities are based solely on the past states, where its performance approaches that of vlmc as the sample size increases. On the other hand, the vlmc often under-estimates the true tree as it fails to identify the covariate effects on the transition probabilities.

\section{Data Analysis: The Hang Seng Index}\label{sec:data}

The Hang Seng Index (HSI) is one of the most important stock market indices in the world. It is an indicator of the Hong Kong overall market performance including commerce and industry, finance, utilities, and properties. As a freefloat market-capitalization-weighted index, it encompasses daily changes of stocks of the largest companies in Hong Kong and as such contains a great deal of information about the overall Asian markets.

Several authors have modeled trends in the HSI using diverse techniques, including for instance time series models (\cite{TongYeung1991}, \cite{BurneckiEtAl2011}, \cite{ChengWang2014}, \cite{Zhang2016}), extreme value theory (\cite{Samuel2008}), multivariate stochastic volatility (\cite{AsaiMcAleer2006}), neural networks (\cite{Kumar2009}), among others. However, in the current globalized era, stock markets are known to be highly influenced by other markets, and thus, one might want to use available information in other markets to predict trends. 
Classical methods in time series and stochastic processes only use the information in the history of the process itself to forecast future observations, while regression models that utilize exogenous covariates as predictors of the future observations may miss the information its own context history may contain.
The proposed methodology bridges this gap by using both the past history of the Hang Seng Index and the information available in other stock market trends.

In this analysis, we model the gains (1) and losses (0) of the HSI as a Variable Length Markov Chain considering as covariates the log returns of three large markets that are known to have correlation with the HSI (\cite{MalliarisUrrutia1992}, \cite{GirardinLiu2007}, \cite{SrikanthAparna2012}), namely the New York Composite Index (NYCI), 
the Nasdaq, and the Nikkei 225 Index. The dataset consists of the daily observations of the HSI, as well as the NYCI, Nasdaq, and Nikkei from May 1998 to June 2019, corresponding to 5395 observations, which are available at  http://finance.yahoo.com. 
Due to the different time zones, the Asian stock markets mostly close before the U.S. stock markets are even open. For this reason, the proposed model, which uses the covariate values of the previous time period, is reasonable since the NYCI log return value at time $t-1$ will be observed after the HSI at time $t-1$, but before the HSI at time $t$. 
In the analysis that follows, similarly to the simulation study, the tuning parameters $s$ and $\gamma$ were chosen to minimize the BIC criterium.
The estimated beta-context is

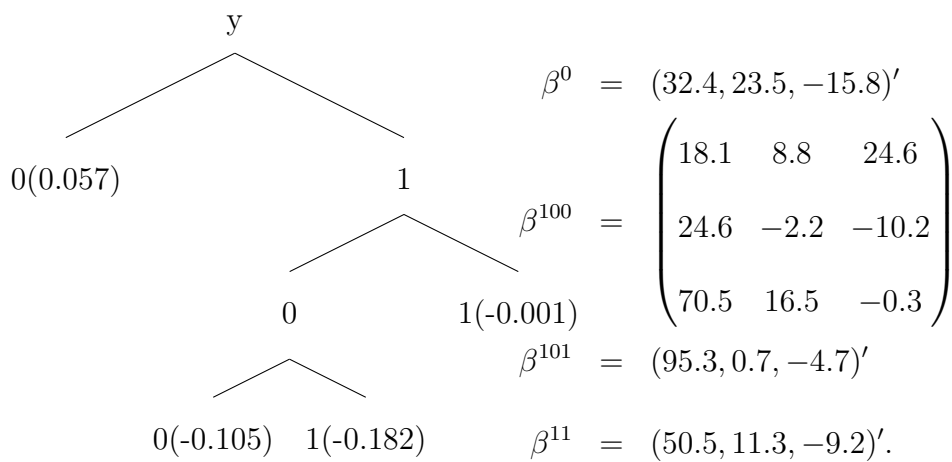
\begin{minipage}{0.45\textwidth}
\Tree[.y [.0(0.057) ]
             [.1 [.0 
                      [.0(-0.105) ]
                      [.1(-0.182) ] ]
                  [.1(-0.001) ]]]
\end{minipage}%
\hfill
\begin{minipage}{0.45\textwidth}
\bqa
\beta^{0} &=& (32.4, 23.5, -15.8)'\\
\beta^{100} &=& \begin{pmatrix} 
18.1 & 8.8 &  24.6  \\
24.6 & -2.2 &  -10.2 \\
70.5 & 16.5 & -0.3 
\end{pmatrix}\\
\beta^{101} &=& (95.3, 0.7, -4.7)'\\
\beta^{11} &=& (50.5, 11.3,  -9.2)'.
\eqa
\end{minipage}

The estimated context tree yields useful insight on the trend dependence structure Hang Seng Index in terms of the NYCI, Nasdaq, and Nikkei indices. The estimated tree suggests that a loss in the previous day composes a context, that is, the next day's gain or loss in the HSI given that a loss occurred in the previous day is independent of the remainder of the history. However, a gain in the HSI in the previous day does not provide sufficient information about the future. Two successive gains are needed to constitute a context, which suggests that two days with positive results already determine the mood of the markets without the need to revisit anything prior to those days. Moreover, a positive result followed by a negative result does not seem to be enough information about what will happen next, so that a third order Markov Chain is required in this case. The signs of the estimated intercepts suggest that, after considering the covariates, if a loss (gain) occurred in the previous day, the process is slightly more (less) likely to have a gain the following day. The estimated covariate parameters indicate that the effect of the Nikkei is mostly negative when NYCI and Nasdaq are in the model, probably as a correction factor.




\markboth{\hfill{\footnotesize\rm Adriano Zanin Zambom  AND Seonjin Kim AND Nancy Lopes Garcia} \hfill}
{\hfill {\footnotesize\rm VLMC with Exogenous Covariates} \hfill}

\bibhang=1.7pc
\bibsep=2pt
\fontsize{9}{14pt plus.8pt minus .6pt}\selectfont
\renewcommand\bibname{\large \bf References}
\expandafter\ifx\csname
natexlab\endcsname\relax\def\natexlab#1{#1}\fi
\expandafter\ifx\csname url\endcsname\relax
  \def\url#1{\texttt{#1}}\fi
\expandafter\ifx\csname urlprefix\endcsname\relax\def\urlprefix{URL}\fi

\lhead[\footnotesize\thepage\fancyplain{}\leftmark]{}\rhead[]{\fancyplain{}\rightmark\footnotesize{} }


%

\vskip .65cm
\noindent
California State University Northridge
\vskip 2pt
\noindent
E-mail: (adriano.zambom@csun.edu)
\vskip 2pt

\noindent
Miami University Ohio
\vskip 2pt
\noindent
E-mail: (kims20@miamioh.edu)

\noindent
State University of Campinas
\vskip 2pt
\noindent
E-mail: (nancyg@unicamp.br)
\end{document}



\renewcommand{\baselinestretch}{2}

\markright{ \hbox{\footnotesize\rm Statistica Sinica: Supplement
}\hfill\\[-13pt]
\hbox{\footnotesize\rm
}\hfill }

\markboth{\hfill{\footnotesize\rm Adriano Zanin Zambom AND Seonjin Kim AND Nancy Lopes Garcia} \hfill}
{\hfill {\footnotesize\rm VLMC with Exogenous Covariates} \hfill}

\renewcommand{\thefootnote}{}
$\ $\par


\fontsize{12}{14pt plus.8pt minus .6pt}\selectfont \vspace{0.8pc}
\centerline{\large\bf Variable Length Markov Chain}
\vspace{2pt} \centerline{\large\bf with Exogenous Covariates}
\vspace{.4cm} \centerline{Adriano Zanin Zambom, Seonjin Kim, and Nancy Lopes Garcia} \vspace{.4cm} \centerline{\it
California State University Northridge, Miami University Ohio,} \centerline{\it and State University of Campinas} \vspace{.55cm} \fontsize{9}{11.5pt plus.8pt minus
.6pt}
\vspace{.55cm}
 \centerline{\bf Supplementary Material}
\selectfont


\def\thefigure{\arabic{figure}}
\def\thetable{\arabic{table}}

\renewcommand{\theequation}{\thesection.\arabic{equation}}

\fontsize{12}{14pt plus.8pt minus .6pt}\selectfont

\setcounter{section}{0} 
\setcounter{equation}{0} 
\def\theequation{S\arabic{section}.\arabic{equation}}
\def\thesection{S\arabic{section}}

\fontsize{12}{14pt plus.8pt minus .6pt}\selectfont


\section{Proofs of Lemma 1 and Theorem 1}

{\bf Proof of Lemma 1}

\begin{proof}
The p-value $\pi^{u}_{-r+1} = 1 - \Psi_{d}(\lambda^{u}_{-r+1})$ tests the hypothesis $H_0^u: \vbeta_{-r+1}^u = 0$, that is, it tests the significance of the coefficient corresponding to the last step ($-r+1$) in the $u$ context's past. 
We use arguments similar to those in the proof of the Wilks Theorem in \cite{Ferguson1996} to establish the result of this Lemma.
In what follows we drop the superscript $u$ in the likelihood function for ease of notation and moreover, we write $\Theta  := \Theta_\tau = \{\vtheta^{u}, \forall u \in \tau\}$ as the vector of all parameters in the parameter tree $\tau_{\vtheta}$, and $\Theta_0$ the parameter vector that satisfies $H_0^u$. Denote
\bqa
l_n(\tilde{\Theta}) &=& \log L\left(\tilde\tau_{\theta}^u; y_1^n, \vx_1^n\right), \text{ and }\\
l_n(\hat\Theta) &=& \log L\left(\hat\tau_{\theta}; y_1^n, \vx_1^n\right),
\eqa
where $\hat\Theta$ is the maximum likelihood estimator, and   $\tilde{\Theta} := \tilde{\Theta}_u$ is estimated parameter vector under the null hypothesis as in equation (2.4) of the paper.
Hence we can write $\lambda^{u}_{-r+1} = 2(l_n(\hat\Theta) - l_n(\tilde\Theta))$. Expanding $l_n(\tilde\Theta)$ about $\hat\Theta$ yields
\bqa
l_n(\tilde\Theta) = l_n(\hat\Theta) + l_n'(\hat\Theta)(\tilde\Theta-\hat\Theta) - n(\tilde\Theta-\hat\Theta)^TI_n(\tilde\Theta)(\tilde\Theta-\hat\Theta),
\eqa
where
\bqa
I_n(\tilde\Theta) = -\frac{1}{n}\int_0^1\int_0^1vl''_n(\hat\Theta + uv(\tilde\Theta-\hat\Theta))dudv. 
\eqa
 Noting that $l_n'(\hat\Theta) = 0$, we have

\bqan\label{test_stat}
\lambda^{u}_{-|\vbeta^u|+1} =  2\sqrt{n}(\tilde\Theta-\hat\Theta)^TI_n(\tilde\Theta)\sqrt{n}(\tilde\Theta-\hat\Theta).
\eqan
Expand $\frac{1}{\sqrt{n}}l_n'(\tilde\Theta)$ about $\hat\Theta$ to obtain
\bqa
\frac{1}{\sqrt{n}}l_n'(\tilde\Theta) = \frac{1}{\sqrt{n}}l_n'(\hat\Theta) + \frac{1}{n}\int_0^1 l''_n(\hat\Theta + v(\tilde\Theta-\hat\Theta))dv\sqrt{n}(\tilde\Theta - \hat\Theta).
\eqa
Denote $J_n(\tilde\Theta) := \frac{1}{n}\int_0^1 l''_n(\hat\Theta + v(\tilde\Theta-\hat\Theta))dv$, 
and write
\bqa
\sqrt{n}(\tilde\Theta - \hat\Theta) = \frac{1}{\sqrt{n}}[J_n(\tilde\Theta)]^{-1}l_n'(\tilde\Theta).
\eqa
Replacing $\sqrt{n}(\tilde\Theta - \hat\Theta)$ in (\ref{test_stat}) yields
\bqa
\lambda^{u}_{-r+1} &=&   2\left[\frac{1}{\sqrt{n}}[J_n(\tilde\Theta)]^{-1}l_n'(\tilde\Theta)\right]^TI_n(\tilde\Theta)\left[\frac{1}{\sqrt{n}}[J_n(\tilde\Theta)]^{-1}l_n'(\tilde\Theta)\right]\\
&=& 2\frac{1}{\sqrt{n}}l_n'(\tilde\Theta)^T\vH\frac{1}{\sqrt{n}}l_n'(\tilde\Theta),
\eqa
where $\vH = [J_n(\tilde\Theta)]^{-1}]^T I_n(\tilde\Theta)[J_n(\tilde\Theta)]^{-1}]$.

Recall that for each $u \in \tau$, the associated parameter vector $\vbeta^u$ has $r$ sets of $d$ parameters different from 0, so that the null hypothesis $H_0^u: \vbeta_{-r+1}^u = 0$ should to be rejected (unless the context has no associated parameter vector, which means $r = 0$). Recall this vector of parameters is denoted by $\Theta$. To find the asymptotic distribution of $l_n'(\tilde\Theta)$ expand about $\Theta$ obtaining
\bqa
\frac{1}{\sqrt{n}}l_n'(\tilde\Theta) &=& \frac{1}{\sqrt{n}}l_n'(\Theta) + \frac{1}{n}\int_0^1 l''_n(\Theta + v(\tilde\Theta-\Theta))dv\sqrt{n}(\tilde\Theta - \Theta)\\
&=& \frac{1}{\sqrt{n}}l_n'(\Theta) + J_n(\tilde\Theta)\sqrt{n}(\tilde\Theta - \Theta).
\eqa

Under the fact that $\Theta$ is the true parameter vector, by the CLT for $\ell$-dependent and non-identically distributed random variables, $\frac{1}{\sqrt{n}}l_n'(\Theta) \dkonv N(0,\mathcal{I}_a(\Theta))$, for some covariance matrix $\mathcal{I}_a(\Theta)$. 
Then, letting $\delta=\tilde\Theta-\Theta_0$ and  $\gamma=\Theta_0-\Theta$, write
\bqa
\lambda^{u}_{-|\vbeta^u|+1} &=&    2\left[\frac{1}{\sqrt{n}}l_n'(\Theta) + J_n(\tilde\Theta)\sqrt{n}(\tilde\Theta - \Theta)\right]^T\vH\left[\frac{1}{\sqrt{n}}l_n'(\Theta) + J_n(\tilde\Theta)\sqrt{n}(\tilde\Theta - \Theta)\right]\\
&=& 2\left[\frac{1}{\sqrt{n}}l_n'(\Theta)\right]^T\vH\left[\frac{1}{\sqrt{n}}l_n'(\Theta)\right] + 4\left[\frac{1}{\sqrt{n}}l_n'(\Theta)\right]^T\vH\left[J_n(\tilde\Theta)\sqrt{n}(\tilde\Theta - \Theta)\right]\\
&& + 2 \left[J_n(\tilde\Theta)\sqrt{n}(\tilde\Theta - \Theta)\right]^T\vH \left[J_n(\tilde\Theta)\sqrt{n}(\tilde\Theta - \Theta)\right]\\
 &=&2\left[\frac{1}{\sqrt{n}}l_n'(\Theta)\right]^T\vH\left[\frac{1}{\sqrt{n}}l_n'(\Theta)\right] + 4\left[\frac{1}{\sqrt{n}}l_n'(\Theta)\right]^T\vH\left[J_n(\tilde\Theta)\sqrt{n}(\delta+\gamma)\right]\\
&& + 2 \left[J_n(\tilde\Theta)\sqrt{n}(\delta+\gamma)\right]^T\vH \left[J_n(\tilde\Theta)\sqrt{n}(\delta+\gamma)\right]\\
 &=&2\left[\frac{1}{\sqrt{n}}l_n'(\Theta)\right]^T\vH\left[\frac{1}{\sqrt{n}}l_n'(\Theta)\right] + 4\left[\frac{1}{\sqrt{n}}l_n'(\Theta)\right]^T\vH\left[J_n(\tilde\Theta)\sqrt{n}\delta\right] \\
&&  + 4\left[\frac{1}{\sqrt{n}}l_n'(\Theta)\right]^T\vH\left[J_n(\tilde\Theta)\sqrt{n}\gamma\right]
 + 2 \left[J_n(\tilde\Theta)\sqrt{n}\delta\right]^T\vH \left[J_n(\tilde\Theta)\sqrt{n}\delta\right]\\
 && +2 \left[J_n(\tilde\Theta)\sqrt{n}\gamma\right]^T\vH \left[J_n(\tilde\Theta)\sqrt{n}\delta\right] +2 \left[J_n(\tilde\Theta)\sqrt{n}\gamma\right]^T\vH \left[J_n(\tilde\Theta)\sqrt{n}\gamma\right]\\
 &=& A + B + C + D + E + F.
\eqa
Note that $\sqrt{n}\delta \to N(0,\sigma_\delta)$ for a constant $\sigma_\delta$, and $\gamma$ is a constant vector with some terms different from zero. 
Note  that  $A, B, D$ are greater or equal to $O_p(1)$ since $\vH \geq O_p(1)$,  $F \geq O_p(n)$, and $C$ and $E$ are greater or equal to $O_p(\sqrt{n})$.
Hence
\bqa
\lambda^{u}_{-r+1} \geq O_p(n) .
\eqa

\end{proof}

{\bf Proof of Theorem 1}
\begin{proof}
a) 

Define
\bqa
U &=& \Big\{  \text{ there exists } w \in \hat\tau \text{ with } wu \in \tau \text{ and } wu \notin \hat\tau^f (u \in \mathcal{Y}^k)\Big\} \text{ and}\\
O &=& \Big\{  \text{ there exists } wu \in \hat\tau (u \in \mathcal{Y}^\infty) \text{ with } w \in \tau^f \text{ and } wu \notin \tau^f \Big\}.
\eqa

Consider $U$. Because we only cut a context $u$ together with its sibling (whole branch), we have
\bqa
P(U) &\leq& \sum_{\substack{u_1 = uw, u_2=uw' \in \tau^f\\ u_1\wr u_2}} P\Big(\left\{\pi^{u_1}_{-|u_1|+1} > \gamma_n\right\}\cap\left\{\pi^{u_2}_{-|u_2|+1} > \gamma_n\right\}\\
&&\hspace{4cm}\cap\{\pi^{u_1u_2} \geq \gamma_n|\vbeta_{-|u_1|+1}^{u_1} = 0, \vbeta_{-|u_2|+1}^{u_2} = 0\}\Big) \\
&\leq& \sum_{\substack{u_1 = uw, u_2=uw' \in \tau^f\\ u_1\wr u_2}} P\Big(\left\{\pi^{u_1}_{-|u_1|+1} > \gamma_n\right\}\Big) 
= P(O_p(n) < \Psi_{d}^{-1}(1-\gamma_n)), 
\eqa
where $\pi^{u_1}_{-|u_1|+1}$ is the p-value for testing the coefficient corresponding to the covariate at step $-|u|+1$ for context $u$, as in equation (2.3) in the paper. By Lemma 1, $P(U)$ goes to zero with the choice of $\gamma_n$ meeting condition C2 since $|\tau|$ is fixed.

We bound $P(O)$ with the sum of the probabilities of overfitting writing
\bqan\label{P_O_Thm2}
P(O) &\leq& \sum_{\substack{u_1\wr u_2 \in \hat{\tau}\\ u_1 = wu, u_2 = wv, w \in \tau^f; u, v \in \mathcal{Y}^\infty}} P\left(\pi^{u_1}_{|u_1|} < \gamma_n\right) + P\left(\pi^{u_2}_{|u_2|} < \gamma_n\right) \nonumber\\
&& \hspace{1cm}+ P(\text{dont cut siblings replacing by parent even after} \nonumber\\
&&  \hspace{2cm}\text{both betas have been reduced})\nonumber\\
 &=&   O\left(\sum_{i=1}^r 2^i\gamma_n\right) 
= O\left(\sum_{i=1}^r 2^i\gamma_n\right)  = O(\gamma_n2^{r+1}) \nonumber\\
&=& O(\gamma_n 2^{log(n)}) = O(\gamma_n n)  = o(1),
\eqan
for the choice of $\gamma_n$ meeting condition C1.

b) 

Define
\bqa
U &=& \Big\{  \text{ there exists } u \in \tau \text{ with } \hat\vtheta^u \in \hat{\tau}_\theta \text{ and } \vtheta^{u} \in \tau_\theta \text{ such that } |\hat\vbeta^u| < |\vbeta^u|\Big\} \text{ and}\\
O &=& \Big\{ \text{ there exists } u \in \tau \text{ with } \hat\vtheta^u \in \hat{\tau}_\theta \text{ and } \vtheta^{u} \in \tau_\theta \text{ such that } |\hat\vbeta^u| > |\vbeta^u| \Big\},
\eqa
where $|\vbeta^u|$ is the length of $\vbeta^u$ as in Definition 2. Then we can write
\bqa
&&P[\hat{\vtheta}^u = \vtheta^u, \forall u \in \tau]^c = \\
&& P(U \cup O \cup \left\{\exists u \in \tau \text{ with } \hat\vtheta^u \in \hat\tau_\theta \text{ and } \vtheta^u \in \tau_\theta \text{ such that } \hat\vbeta^u \neq \vbeta^u \cap |\hat\vbeta^u| = |\vbeta^u|\right\})
\eqa

Consider $U$.
We have
\bqan \label{U_n}
P(U) &\leq& \sum_{u \in \tau} P\left(\pi^{u}_{-|\vbeta^u|+1} > \gamma_n\right) = \sum_{u \in \tau} P\left(1 - \Psi_{d}\left(\lambda^{u}_{-|\vbeta^u|+1}\right) > \gamma_n\right)\nonumber\\
&&= \sum_{u \in \tau} P\left(\lambda^{u}_{-|\vbeta^u|+1} < \Psi_{d}^{-1}(1-\gamma_n)\right),
\eqan
where $\pi^{u}_{-|\vbeta^u|+1} = 1 - \Psi_{d}\left(\lambda^{u}_{-|\vbeta^u|+1}\right)$, and $\lambda^{u}_{-|\vbeta^u|+1}$ is defined as in equation (2.3) in the paper. 
Hence
\bqa
P(\lambda^{u}_{|\vbeta^u|} < \Psi_{d}^{-1}(1-\gamma_n)) &\leq& P(O_p(n) < \Psi_{d}^{-1}(1-\gamma_n)).
\eqa

For this probability to go to 0 it suffices to choose $\gamma_n$ such that $\Psi_{d}^{-1}(1-\gamma_n) = o(n)$.
By Inglot (),
$\Psi^{-1}_d(1-\gamma_n) \leq d + 2log(1/\gamma_n) + 2\sqrt{dlog(1/\gamma_n)}$, so that the choice of $\gamma_n$ has to be such that $(1/n)\log(1/\gamma_n) \to 0$, which is guaranteed by condition C2.

Hence
\bqa
P(U) &\leq&  \sum_{u \in \tau} P\left(\lambda^{u}_{-|\vbeta^u|+1} < \Psi_{d}^{-1}(1-\gamma_n)\right) = |\tau|o(1) = o(1).
\eqa

Now consider $O$. Overfitting in this case is due to a rejection of a hypothesis test for a non-significant coefficient, that is
\bqan\label{P_O_Thm1}
P(O) &\leq& \sum_{j > |\vbeta^u|, u \in \tau^f} P(\pi^{u}_{-j+1} < \gamma_n) = \sum_{j > |\vbeta^u|, u \in \tau} \gamma_n\nonumber\\
 &&\hspace{-1cm}= \gamma_n\sum_{i=1}^r2^i = O(\gamma_n2^{r+1}) = O(\gamma_n 2^{log(n)}) = O(\gamma_n n)  = o(1),
\eqan
by condition C1, where $\pi^{u}_{-j+1}$ is the p-value for $H_0^u: \vbeta_{-j+1}^u = 0$, and $r$ is the order of the maximal tree $\tau_{max}$, which by condition C3 is $r = O(log(n))$.

c) The result in part c) of the theorem  follows from the fact that $\hat\Theta^u$ is a maximum likelihood estimator, which is  consistent for $\Theta$.

\end{proof}
